\documentstyle[psfig,12pt]{article}

\def\bold#1{\setbox0=\hbox{$#1$}%
      \kern-.025em\copy0\kern-\wd0
      \kern.05em\copy0\kern-\wd0
      \kern-.025em\raise.0433em\box0 }
   
\def\apleq{\begin{array}[c]{c}{\scriptstyle <}\\[-4.mm]{\scriptstyle \sim} \end{array}}
\def\apgeq{\begin{array}[c]{c}{\scriptstyle >}\\[-4.mm]{\scriptstyle \sim} \end{array}}

\newcommand{\beqa}{\begin{eqnarray}}
\newcommand{\eeqa}{\end{eqnarray}}
\newcommand{\non}{\nonumber}

\begin{document}
\renewcommand{\thefootnote}{\fnsymbol{footnote}}

\begin{titlepage}

\hfill TAN-FNT-97-05
 
\begin{center}
\vspace*{2.0cm}
{\large\bf BABY SKYRMIONS ON THE SPHERE}
\vskip 1.5cm

{N.N. Scoccola
\footnote[1]{e-mail: scoccola@tandar.cnea.edu.ar}
$^\ddagger$ 
and 
D.R Bes
\footnote[2]{e-mail: bes@tandar.cnea.edu.ar}
\footnote[3]{Fellow of the CONICET, Buenos Aires, Argentina.
}}

\vskip .2cm
{\it Physics Department, Comisi\'on Nacional de Energ\'{\i}a At\'omica,
          Av.Libertador 8250, (1429) Buenos Aires, Argentina.}

\vskip 2.cm

August 1997

\vskip 2.cm
{\bf ABSTRACT}\\
\begin{quotation}
We study a model for two-dimensional skyrmions on 
a sphere of radius $L$. Such model simulates a 
skyrmion lattice of density $W/(2 \pi L^2)$,
where $W$ is the skyrmion winding number.  
We show that, to a very good approximation, physical results depend 
only on the product $\alpha L^4$, where $\alpha$ is the strength of 
potential term. 
In the range $\alpha L^4  \apleq 3$ the order 
parameter vanishes, there is 
a uniform distribution of the density over the whole surface
and the energy of the $W=2$ sector lies
above twice the energy of the $W=1$ sector. 
If $\alpha L^4 \apgeq 6$ the order parameter approaches unity
and the density concentrates near one of the poles. Moreover
the disoliton is always bound. We also present a variational
solution to the field equations for which the pure $\alpha L^4-$dependence
is exact. Finally, some consequences of our results for the Quantum
Hall Effect are discussed.

\end{quotation}
\end{center}

PACS number(s): 11.10.Lm, 11.27.+d, 73.40.Hm


\end{titlepage}

\renewcommand{\thefootnote}{\arabic{footnote}}

Although skyrmions were originally introduced to describe baryons
in the context of non-linear meson theories \cite{Sky61}, they are playing
an increasing role in other areas of physics as well.
In the present paper we consider the case of ``baby skyrmions'', 
i.e. skyrmions in two spatial dimensions.  
For example, 2D skyrmions are used to model the bubbles that appear in certain 
condense matter systems in the presence of an external magnetic field ; they could 
provide a mechanism associated with the disappearance of antiferromagnetism and 
the onset of $H T_c$ superconductivity; etc.. Recently 
it has been suggested \cite{SKKR93}
that they may be responsible for many features of the Quantum Hall Effect (QHE).
Baby skyrmions have been also studied in the context of
strong interactions as a toy-model in order to understand the more complicated
dynamics of usual skyrmions which live, of course, in 3 spatial 
dimensions \cite{PSZ95}.

For many of these applications it is of great interest to
consider not only the properties of few baby skyrmion systems but
also the behavior of such properties as the density of 
skyrmions increases. So far, this problem has been studied by putting
baby skyrmions on a lattice, which is, of course, a rather
cumbersome numerical task (see e.g. Ref. \cite{BFCMD95}). Analogous 
studies have been done for 3D skyrmions to understand baryon properties in dense
matter and the appearance of phase transitions such as
chiral restoration, deconfinement, etc \cite{Kle85}. However, in the 3D case
it was  realized that many of the qualitative results
 of the lattice calculations (and even some quantitative ones)
could be obtained in a much easier way by studying the behavior 
of few skyrmions on a hypersphere \cite{JWC88}. 
This approach was introduced in Ref. \cite{MR86} and proposes to replace
the periodic array in flat space ${\bf R}^3$ by a one (or few) skyrmion system
on the compact manifold ${\bf S}^3(L)$ of radius $L$. The finite baryon number
on ${\bf S}^3(L)$ corresponds to a finite baryon density, so that one obtains
a model for skyrmion matter which is much simpler to study than any lattice
model. The density of this matter can be increased by decreasing $L$.
The aim of the present paper is to 
extend these ideas to the case of 2D skyrmions on the sphere.

Baby skyrmions are obtained as non-trivial solutions of the well-known non-linear O(3) model.
This model consists of three real scalar fields $\phi_a$ ($a=1,2,3$) subject to the constraint
${\vec \phi}\cdot{\vec \phi}=1$. The equations of motion admit static solutions with finite energy which represent 
a mapping of ${\bf R}^2_{spat}$ into $S^2_{int}$. They are characterized by the 
winding number $W$ and the density $\rho$
 
\beqa
W=\frac{1}{8 \pi}\int d^2 r \,\rho \qquad \qquad ; 
\qquad \qquad \rho \equiv  \epsilon_{i j} 
\ \vec \phi \cdot \left( \partial^i \vec \phi \times \partial^j \vec \phi \right) 
\eeqa

\noindent
where $i,j$ stand for the spatial coordinates.
We start from the energy functional   
\beqa
E &=& E^{(2)} + E^{(4)} + E^{(p)}\non \\[7.mm]
E^{(2)} = \frac{1}{2} \int d^2r \ \partial_{i} {\vec \phi} 
\cdot 
\partial^{i} {\vec \phi}
\ \ ; \ \
E^{(4)} &=& \frac{1}{8}\int d^2r \ \rho \ \rho \ \ ; \ \
E^{(p)} = {\alpha\over2} \int d^2r \ (\hat n_3 - 
{\vec \phi})^2 \label{lag0}
\eeqa

\noindent
where $\hat n_3$ is a unit vector in the third direction in internal space and
$\alpha$ is a parameter that is assumed  positive. This type of energy 
functional is obtained as the static limit of either the relativistic $O(3)$ 
model or the non-relativistic Chern-Simons theory.

In 2D the quadratic term in the field derivatives gives a 
scale independent contribution to the soliton mass, due to 
the well known compensation between the derivatives and
the integration \cite{BP75}. Thus, the stabilization procedure for baby  skyrmions 
is usually based on  a competition between a quartic term and a linear potential term.
This last (attractive) interaction tends to concentrate the skyrmion at 
$\phi_3=1$, yields a contribution to the total energy proportional to the
square of the soliton size $\lambda$ and destroys the original O(3) symmetry.
In condense matter physics this potential is usually related to the
magnetic interactions (e.g. Zeeman term in QHE). 
In the framework of hadron physics it would correspond to the 
explicit chiral symmetry breaking driven by the pion mass term and therefore 
considered to be only a small perturbation, and often neglected. 
The collapse of the field is prevented by means of the
quartic term  which yields a contribution proportional to $1/\lambda^2$.
In QHE this contact interaction would be a ``poor man" mocking of the 
Coulomb repulsion, while in hadron physics it can be understood as a large mass limit
of the $\rho$-meson exchange. It should be mentioned that, as in the 3D case, the quartic
term can be replaced by higher order derivative 
terms\footnote{In fact, using 
particular combinations of derivative terms it is possible even to stabilize the soliton 
in the absence of the potential term. Although interesting, we
do not consider this possibility here. For details, 
see Ref.\cite{JVZC88}.}. From the 
experience in 3D we do not expect major changes in the behavior of most 
of the quantities to be discussed below.

The simple relation between the coefficients of the quadratic and quartic terms 
in (\ref{lag0}) is obtained through a convenient choice of the unit of length.
Therefore, in flat space there is only one parameter in the model, namely the strength $\alpha$ 
of the potential term, aside from an overall multiplying factor which is already omitted in 
(\ref{lag0}). 

It is interesting to note that using the inequality
\beqa
0&\leq &\int d^2r \,\left[
\frac{1}{4}\left(\partial_{i}{\vec \phi}\pm
\epsilon_{i j}{\vec \phi} \times \partial^{j}{\vec \phi} \right)^2
+ \frac{1}{2} \left(\frac{1}{2} \epsilon_{i 
j} \partial^{i}{\vec \phi} \times \partial^{j}{\vec \phi}
\pm \sqrt{\alpha}({\hat n}_3-{\vec \phi})\right)^2 \right] 
\eeqa
one can find the Bogomol'nyi bound for the baby skyrmion energy. One obtains 
\beqa
E \ge 4\pi k \ \left(1+\sqrt{\alpha} \right) 
\label{bound}
\eeqa
which improves over the usual bound \cite{PSZ95} through the 
contribution proportional to $\sqrt{\alpha}$.

As mentioned above, our aim is to extend the model above by going from
${\bf R}^2_{sp}$ to ${\bf S}^2_{sp}(L)$ where $L$ is the radius of the two-sphere.
A convenient choice of coordinates on ${\bf S}^2_{sp}(L)$ is given by the 
conventional polar coordinates $\theta, \varphi$ ($0\leq \theta \leq \pi$
and $0\leq \varphi \leq 2\pi$) 
\beqa
x=L \sin{\theta} \cos{\varphi}\;;\;\;\;\;
y=L \sin{\theta} \sin{\varphi}
\eeqa
The jacobian of the transformation and the metric associated with the polar coordinates 
are
\beqa
J    &=&-L^2 \sin{\theta} \non \\
ds^2 &= & L^2 (\sin^2{\theta} d\varphi^2 +d\theta^2) \label{metric}
\eeqa

In order to obtain explicit static solutions in the winding number $W=k$ sector 
we introduce the hedgehog parameterization
\beqa
\phi_1=\sin{f}\,\cos{k \varphi}\;;\;\;\;\;
\phi_2=\sin{f}\,\sin{k \varphi}\;;\;\;\;\;
\phi_3=\cos{f} \label{hedge}
\eeqa
where the ``radial'' profile $f=f(\theta)$ is subject to the boundary
conditions $f(0)=\pi$ and $f(\pi)=0$.   

          Using (\ref{hedge}) and (\ref{metric}) we may 
express $E_k$, the  soliton energy divided by $4 \pi k$,
 in terms of the
profile $f$ as
\beqa
E_k=\frac{1}{4k}\int d\theta \,\sin{\theta}\,
\left[f'^2+k^2 \left( \sin f\over\sin \theta\right)^2 
+ \frac{k^2}{L^2}f'^2\left( \sin f\over\sin \theta\right)^2+
2 \alpha L^2 (1 - \cos{f})\right] \label{mass}
\eeqa
while the winding number density results
\beqa
\rho_k &=&-{2 k\over{L^2}}  \ f' \ { \sin f\over \sin \theta} \label{wden}
\eeqa
In Eqs.(\ref{mass},\ref{wden}) a prime denotes differentiation with respect to $\theta$.

Minimizing $E_k$ we obtain the eq. of motion for $f$. It results 
\beqa
& & \left[ 1 + \frac{k^2}{L^2} \left( \sin f\over\sin \theta\right)^2 \right] f'' +
\nonumber \\
& & \qquad +
\left[ f'-k^2 {\sin 2 f\over\sin 2 \theta} + \frac{k^2}{L^2} f' {\sin f\over\sin \theta} 
\left(f' {\cos f\over\cos \theta } - {\sin f\over\sin \theta} \right) \right] \cot{\theta} 
 - \alpha L^2 \sin{f}=0
\label{lageq}
\eeqa

Fig. 1(a)  displays the profile function $f(\theta)$
in the $k=1$ sector for different values of $L$, and 
$\alpha=0.1$. Fig. 1(b) shows the same magnitudes but for different values 
of $\alpha$, and $L=2$. 
For very dense systems or for very small values of $\alpha$
the identity map $f=\pi-\theta$ constitutes a valid (although not exact)
solution for $f$.
As shown in Figs. 1(c) and 1(d),  it corresponds to a uniform 
density $\rho_1$ over the surface of the
sphere.
As $L$ or $\alpha$ increases, both  the profile and the density
become more concentrated near $\theta \approx 0$. 

The fact that similar results are obtained by varying 
either $L$ or $\alpha$ raises the question whether the
baby skyrmion on the sphere is a single-parameter  model. The origin  of such uniqueness 
may be traced back to the fairly good independence of the quadratic term
$E^{(2)}$  respective to both $L$ and $\alpha$ . This can be clearly
seen in Figs.2(a) and 2(b). There it is shown the behaviour of the different 
contributions to the soliton energy  as function of $L$ and $\alpha$,
respectively.  As a consequence, the physics of the model is 
determined only by the interplay between the quartic and the 
potential term,  and thus depends basically on the product 
$\alpha L^4$. The fact that some quantities (e.g. the mean square radius 
in Fig.2)
behave differently as function of $L$ or as function of $\alpha$
is simply due to explicit dependencies on $L$.  

It is clear from Fig.2 that we may distinguish two regions
which display  different behaviors of the physical magnitudes.
For large values of $\alpha L^4$ the order parameter
defined as the mean value of $\phi_3$
\begin{equation}
\bar \phi_3 = {1\over2} \int d\theta \sin{\theta} \ \phi_3
\label{ord}
\end{equation}
tends to 1; the quartic and potential contributions
$E^{(4)}<E^{(p)}$ become rather similar;   
the energy of the $k=2$ sector is  smaller than (twice)
the one of the $k=1$ sector;
the mean square radius 
\begin{equation}
{\bar r_k} = \left[ 
2 \pi L^2 \int d\theta \sin{\theta} \ \rho_k \ (\sin\theta)^2 
\right]^{1/2} \label{rmsqt}
\end{equation}
decreases, if plotted in units of $L$.
On the contrary, for small values of
$\alpha L^4$ the order parameter vanishes; $E^{(4)}>>E^{(p)}$;
the energy of the $k=2$ sector is  larger than (twice)
the one of the $k=1$ sector; the mean square radius remains
approximately constant.  
Between the two extreme behaviors there is a  broad transition region 
corresponding approximately  to the values $3 \apleq \alpha L^4 \apleq 6$.
This differs from the situation in the absence of the quadratic term
$E^{(2)}$\cite{GP97}.
 
The behaviour of the physical quantities as function of $L$
is, in general, similar to the one obtained in 3D \cite{JWC88}. 
Due to the presence of the potential term, however, the sharp 
transition found in 3D (in the chiral limit) turns out to be a 
broad one here. Such term is also responsible for the absence 
of ``swelling" of the mean square radius in our model. This seems
to be the only qualitative feature of 3D skyrmions which is
not shared by  baby skyrmions. 

In the context of QHE there is an undergoing discussion
about the stability of the disoliton at low densities, 
with respect to the decay into two $k=1$ solitons.
In Ref. \cite{NK97}, it was found that the disoliton
is stable independently of the strength of the potential 
term. This seems to be in contradiction with  
Ref. \cite{LLKS97}, where it is shown that there is a critical 
strength above which the disolitons become unstable.
Our results indicate that at zero density
the disoliton is always stable. They also confirm the
conjeture made in Ref. \cite{NK97} that, as density
increases (and therefore $L$ decreases), there might be a 
transition above which disolitons become unstable.
The fact that at zero density (flat space) the disoliton
is stable is in agreement with the result
obtained in 3D. There the disoliton is
supposed to represent the deuteron. As well-known, at
classical level, the lowest diskyrmion (the doughnut) is 
rather strongly bound \cite{KS87}. 
Recently \cite{LMS95} it has been
shown that quantum effects tend to decrease such binding.
It is clear that in order to make more solid statements
such quantal effects should be introduced in our model  
through a proper quantization method (see, for instance, 
Ref. \cite{GRKR95}). Another point to take into account
when applying our model to QHE is the long range nature
the Coulomb interaction. As mentioned above a very crude approximation 
has been used in the present work. Namely,
we use a local density-density interaction as a four-derivative
term. The consequences of replacing such term by a long range 
interaction is currently under investigation.

We finish by presenting an approximate variational
solution for the  baby skyrmions on the sphere.
For this purpose one can use the solutions of the
equation of motion (\ref{lageq}) in the absence of quartic
and potential terms. They can be written as
\beqa
\tan{\frac{f}{2}}
=2 \left[ \frac{\eta}{2} \cot{\frac{\theta}{2}} \right]^k \label{fapprox}
\eeqa
where $\eta$ is an arbitrary constant. These
solutions reduce to the known expression in flat space
\cite{BP75}
and saturate the corresponding Bolomol'nyi bound (i.e. Eq.(\ref{bound})
with $\alpha = 0$) for any value of $\eta$. 
Therefore when used as variational solution of the
full model the value of $\eta$ turns out be only a function of the combination 
$\alpha L^4$. For example, minimization
of the soliton energy in the $k=1$ sector leads to the following equation for 
$\eta$
\begin{equation}
\alpha\,L^4 = \frac{(\eta^2-1)^4(\eta^2+1)}
{24 \eta^4[\eta^2-1-(\eta^2+1)\log{\eta}]}
\label{minpar}
\end{equation}
All the features that are present in the exact solution of the hedgehog 
ansatz are qualitatively reproduced, the quantitative  agreement 
lying within few percents. A detailed comparison will be given elsewhere.

To conclude, we have studied  the consequences of locating one or two 
baby skyrmions on the surface of a sphere of radius $L$. Thus
the model covers the whole range between the two limits
of low densities (flat space) and of high density (small $L$). The soliton 
is stabilized through the presence of a (repulsive) quartic
term in the field derivatives and of an (attractive)
potential term proportional to the strength  $\alpha$ of an
external field, in addition to the usual quadratic term.
A value of the Bogomol'nyi bound which is higher than the 
one currently used in the literature on baby skyrmions has been 
obtained.
Within the hedgehog approximation, the model was solved exactly.
Although in principle the model has two independent
parameters, we show that the product $\alpha L^4$ determines
the physically meaningful properties of the system.
The solution displays two characteristic regimes: the order 
parameter vanishes and there is 
a uniform distribution of the density over the whole surface
for the range $\alpha L^4 \apleq  3$ ; on the contrary, the order 
parameter approaches unity
and the density concentrates near $\theta \approx 0$
if $\alpha L^4 \apgeq 6$. In the first region the energy of the 
$k=2$ sector lies
above twice the energy of the $k=1$ sector. The opposite 
is true for the soliton region, in which the disoliton
is {\em always} bound. Finally, a variational solution
that realizes exactly the pure $\alpha L^4-$dependence 
is given.
We believe that some of the results obtained in the 
present work can be directly extended to interesting
issues such as phase transitions in Quantum Hall Effect.
There, the skyrmion density would be related with the
departure of the filling factor from one. 
However, some improvements, such as e.g. inclusion of long range 
interactions, have to be made before drawing stronger conclusions.
We hope to be able to report on these matters in forthcoming
publications.   
   
\vskip 1.2cm

We would like to thank C.L. Schat and G. Zemba for enlightening
discussions. Partial support of Fundaci\'on Antorchas is
gratefully acknowledged.


                 
                 

                 
                 

\begin{figure}
\centerline{\psfig{figure=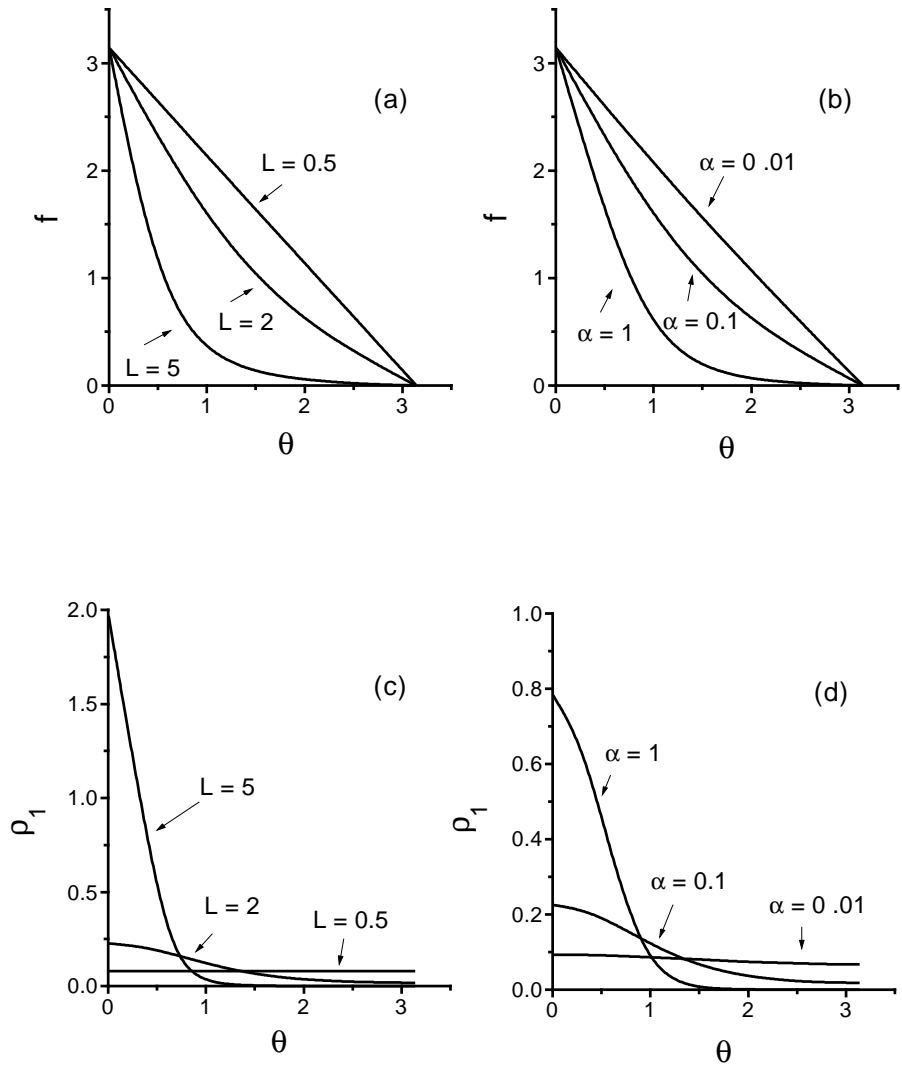,height=20cm}}
\vspace{.25cm}
\protect\caption{\it Soliton profiles (a) and (b) and 
winding number densities (c) and (d). The densities 
(\ref{wden}) are plotted in units of $8 \pi/L^2$.}
\label{fig1}
\end{figure}

\vfill

\pagebreak
\begin{figure}
\centerline{\psfig{figure=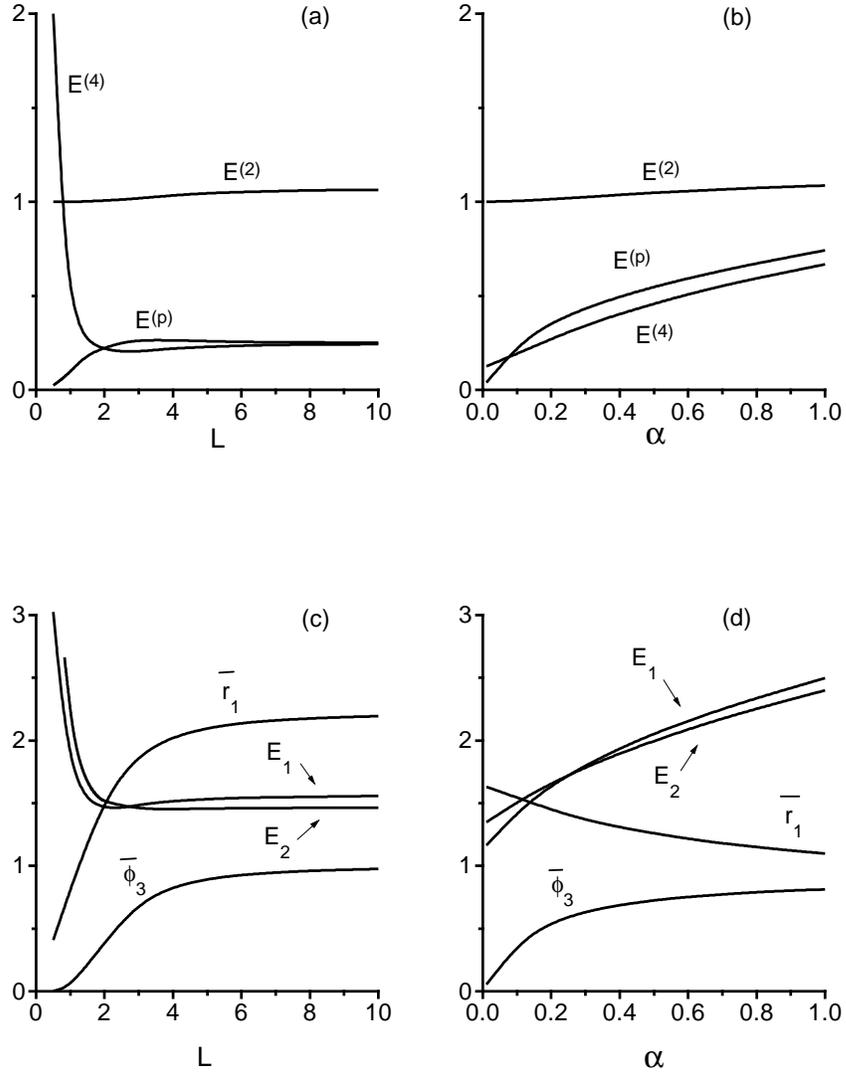,height=20cm}}
\vspace{.25cm}
\protect\caption{\it Various properties of the solution. Figs. (a)
and (b) contain the different contributions (\ref{lag0}) to the 
soliton energy for winding number k=1 (in units of $4 \pi$). Figs.
(c) and (d) include the total soliton energies (\ref{mass}),
the order parameter (\ref{ord}) and the mean square radius
(\ref{rmsqt}).}

\label{fig2}
\end{figure}

\end{document}